# Token-based identity management in the distributed cloud

Ivana Kovacevic[0000-0001-9418-8814]. Tamara Rankovic[0000-0002-5265-4626], Milan Stojkov[0000-0002-0602-0606] and Milos Simic [0000-0001-8646-1569]

[1] Faculty of technical sciences, Novi Sad, Serbia
`kovacevic.ivana@uns.ac.rs`

**Abstract.** The immense shift to cloud computing has brought changes in security and privacy requirements, impacting critical Identity Management (IdM) services. Currently, many IdM systems and solutions are accessible as cloud services, delivering identity services for applications in closed domains and the public cloud. This research paper centers on identity management in distributed environments, emphasizing the importance of robust up to date auhtorization mechanisms. The paper concentrates on implementing robust security paradigms to minimize communication overhead among services while preserving privacy and access control. The key contribution focuses on solving the problem of restricted access to resources in cases when the authentication token is still valid, but permissions are updated. The proposed solution incorporates an Identity and Access Management (IAM) server as a component that authenticates all external requests. The IAM server's key responsibilities include maintaining user data, assigning privileges within the system, and authorization. Furthermore, it empowers users by offering an Application Programming Interface (API) for managing users and their rights within the same organization, providing finer granularity in authorization. The IAM server has been integrated with a configuration dissemination tool designed as a distributed cloud infrastructure to evaluate the solution.

## 1  Introduction

In the modern era, more business solutions are moving towards using the cloud, given that cloud computing aims to improve the management of computing resources by combining concepts such as elasticity, on-demand use, and dynamic resource allocation [1]. This tremendous shift to cloud computing resulted in changes in requirements for security and privacy, affecting critical identity management (IdM) services as well. Currently, numerous IdM systems and solutions are available as cloud services, providing identity services to applications operated in closed domains and the public cloud [2]. Traditional identity management systems rely on centralized identity providers (IdPs) that create, manage, and maintain the identity information of their users or smart devices while providing authentication mechanisms to service providers [3]. This research paper aims to propose one solution for identity management within distributed environments, emphasizing the criticality of robust authentication and the principle of least privilege for executing specific tasks. The general idea of token-based authentication involves the exchange of client authentication credentials, with the server responsible for generating and verifying the authentication token. A significant challenge in token-based authentication arises from the potential reuse of the same token, particularly in cases where the token is not altered or expired, but user privileges have been updated. The main research questions this paper addresseees are as follows:

1. Can an architectural shift of the centralized IAM component be made to the edge of application layer of the cloud to minimize redundant communication between internal services?

2. Can unified token objects be used among services to avoid multiple data formats and ease token verification and user authorization by introducing login and permissions tokens decoupling?

It is argued that systems avoiding security mechanisms or having inadequate security configurations pose an elevated risk, potentially resulting in substantial and irreversible breaches affecting users and sensitive data. The absence of proper security measures jeopardizes the system's integrity and introduces profound vulnerabilities. Moreover, improper security configurations can lead to substantial delays in communication between components within distributed systems. This latency, in turn, has cascading effects on the overall performance of the system, leading to inefficiencies and suboptimal user experience. Numerous requests exchanged between services need a confirmation mechanism to ascertain that the sender is authorized to receive or send specific information. The system must depend on an identity and access management service to facilitate this. This service plays a crucial role in maintaining entities, their organizations, and access rights. Additionally, it ensures proper authentication through the issuance of tamper-proof tokens, enhancing the overall security of the system.

The rest of the paper is organized as follows. Section II describes the related work for this research. Section III describes token-based identity management principles and gives an overview of IAM architecture. Section IV presents the implementation of the IAM server, the authentication and verification flow referring to a use case of configuration dissemination tool that is developed as a module in distributed cloud. Discussion of the proposed solution is represented in section V. Finally, conclusion and future directions are described in section VI.

## 2    Related work

Research has been conducted in both academic and industrial environments regarding centralized identity management. In [4], Jan Kubovy et al. introduced the Central Authentication & Authorization System (CAAS), an implementation of the OpenId standard [5] and the OAuth2.0 [6] framework that uses the token encryption. The concept of additional token encryption stemmed from the recognition that numerous applications still do not depend on Transport Layer Security and its appropriate configuration [7]. The CAAS environment runs at least one authentication server and one authorization server. These servers handle access privileges for arbitrary users and clients to resources of one or more resource servers [4]. The main purpose of authentication server is to manage user data and verify user identity. Another integral component, the authorization server, is tasked with the administration of user permissions, dictating user access to resources. To fulfill this role, it relies on the OAuth 2.0 protocol.

Dhungana, R. D., et al. in [8] proposed the design, deployment, and integration of an identity management framework into the cloud networking (CloNe) infrastructure. The framework is based on UMA protocol and it provides authentication, authorization and identity management of entities in CloNe architecture. Additionally, it enables federated identity management and management of access control policies across different cloud providers [8].

CREDENTIAL [9] consists of multiple services for storing, managing and sharing digital identity and other sensitive information. Security of these services relies on the combination of strong hardware-based multi-factor authentication with end-to-end encryption representing a signifcant advantage over current password-based authentication schemes [9]. The solution proposed by Kostopoulos et al. leverages proxy re-encryption [10] and redactable signatures [11], facilitating secure access to identity data within cloud-based systems.

In [12], authors suggested a model that uses a framework that harnesses the stateless and secure nature of JWT for client authentication and session management. Furthermore, authorized access to protected cloud SaaS resources has been efficiently managed. Accordingly, a Policy Match Gate (PMG) component and a Policy Activity Monitor (PAM) component have been introduced. In addition, other subcomponents, such as a Policy Validation Unit (PVU) and a Policy Proxy DB (PPDB), have also been established for optimized service delivery. What distinguishes our approach from related work is the reduced number of components, leading to less communication between them. Moreover, unlike the solutions mentioned earlier, our approach refrains from storing permissions in the login token, thereby ensuring the non-leakage of sensitive permissions. The maintenance of up-to-date permissions is achieved through the generation of distinct tokens with permissions internally for each request.

## 3     Token-based identity management

Users can be authenticated to access resources via passwords, biometric, token-based or through certificates [13]. Regardless of the authentication mechanism, secure access between users and the cloud must be ensured. To mitigate the risk of increasing response times per request, this solution depends on tokens as a means of enhancing security and ensuring the integrity of users. Tokens are system generated arbitrary construct that asserts the identity of what it claims to be [14]. Token-based authentication involves the exchange of client authentication credentials, with the server undetaking the responsibilities of generating and verifying the authentication token. In subsequent requests, the generated token is incorporated into request headers.

A major obstacle in token-based authentication arises from the potential reuse of the valid token that stores user permissions in cases that permissions were changed in the meantime. In the approach outlined in this paper, this issue is addressed by generating two distinct tokens. The first token is acquired following a successful login, while the second is a one-time authorization token customized for each individual request. According to RFC [15], there are several types of tokens, each characterized by attributes like reusability, renewability, and the expiration. Specifically, the authentication process relies on the utilization of an access_token. This token is designed for multiple uses within its defined validity period, and it cannot be renewed. Its transmission is securely encrypted via HTTPS. On the other hand, for the transmission of granted permissions between services, a novel token is dynamically generated for each request. This token has a very short lifespan, resembling a permishable_token. This deliberate design aims to enhance security by restricting token use and duration, at the same time minimizing the impact of potential permission updates.

### 3.1    IAM infrastructure

IAM server is part of an authentication inrfastructure, wider term covering all services included in the proccess of authentication, as well as trusted third aprty and databases needed for user attributes, organizations, permissions, and relations among them.
Instead of developing authentication methods tailored for each service within the distributed cloud framework, it is argued that a more a pragmatic alternative would be to isolate one authentication infrastructure. This infrastructure can subsequently be adapted for use with both currently established services and those that may be introduced in the future. Such an approach not only favour the enforcement of a consistent authentication policy across the entire system but also ensures its autonomy from the underlying platform. In this architecture, users are assigned a unique set of credentials. The adoption of a single account per user coordinates with a standardized data format and authentication protocol, ensuring compatibility with every entity engaging in communication with the IAM server.

Proposed shift to the edge of the application layer serves as a boundary to the rest of system in case authentication and authorization requirements are fullfilled. To achieve faster responsiveness of the cloud platform, requests are not forwarded if any of the follwing is missing: authentication token, authorization token, proper permissions within authorization token. Simple architecture overview is depicted on Fig. 1, showing the placement of the IAM server as a proxy between API gateway and services that manage namespaces, configurations and nodes inside the constellations module [16][17] in a distributed cloud. This setup serves as the platform for evaluating the proposed solution. Policy engine and authentication service are components of the IAM server and their capabilities, usages and mutual communication will be explained in details in section 4.

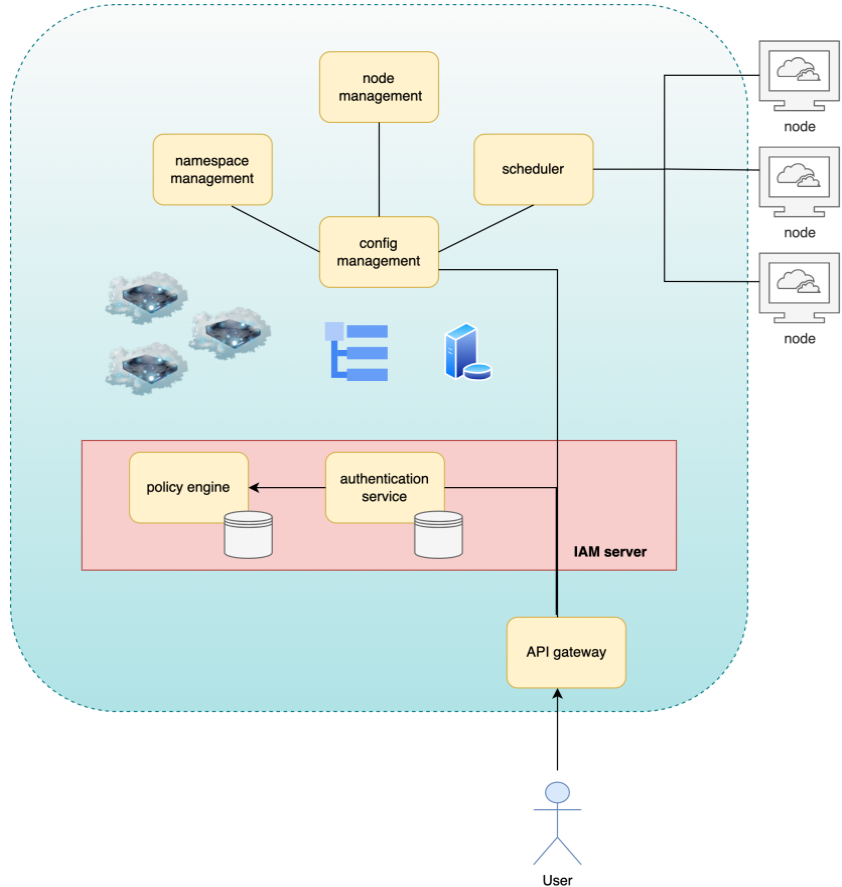

**Fig. 1.** IAM server within the distributed cloud platform

## 4   IAM server implementation

Configurable, highly available identity management services are integrated within configuration dissemination tool within the distributed cloud. This tool is part of the contellations, open-source distributed cloud (DC) platform [17]. The foundational structure of the IAM server relies on the integration of cryptographic mechanisms involving three core components - identity service, authorization service and trusted third-party (TTP) provider.

The process of authentication and access control management is shifted to the edge of an application layer of the system, making the IAM server a proxy for all requests entering the distributed platform. Requests are directed to specific services from the gateway, which handles the routing of requests as well as a protocol conversion for consistent communication among the remaining components of the system. Each request will reach its destination service only if it passes proper authentication and if it is confirmed that the sender has an authority to perofrm desired action.

The identity service bears the responsibility for user account management. Users have the ability to register on the platform, update their credentials, and administer users and permissions within their organization. Upon registration, each user is automatically designated as the owner of their organization, affording a specific level of control and granularity. This empowerment allows users to add or remove members and employ fine-grained control over permissions within the organization.

The authorization service manages relations among users, organizations, and permissions. It expands the Attribute-based access control (ABAC) model for defining policies, incorporating hierarchies of users and objects. This service is also responsible for managing updates in permissions within organizations and communicates directly with the identity service through the remote procedure call (gRPC) [18] framework.

The selection of the trusted third-party provider is based on its lightweight interface and reliability. This provider securely stores user credentials and facilitates password authentication. Due to safety considerations, its secret storage is the exclusive location for storing passwords within this distributed cloud. The solution also leverages a third-party provider for additional security measures such as storing symmetric keys and certificate management. Upon authentication, the user obtains a service expiration token, which cannot be renewed and is exclusively used for user authentication. The provider is responsible for validating this token and communicates with the identity service.

### 4.1   Process of registration

The registration process comprises multiple steps. Initially, the user submits their data in javascript object notation (JSON) format, including a unique username, password, personal information, and the organization's name. The received data undergoes processing and validation. The username must be unique as it is a prerequisite for future logins. If the user omits an organization name, a default organization is created, generated using the chosen username. The newly registered user is then selected as the owner of his organization. To prevent malicious users from assigning themselves as owners of existing organizations, which could compromise access rights and lead to the misuse of member management functionalities, organizations must also be unique within the system. Consequently, registering as the owner of an already existing organization is strictly forbidden.

Additional steps to complete the registration process include: (1) The user is successfully registered on the trusted third-party platform with their username and password. No additional data is sent to the TTP provider, (2) A new organization is stored in a distributed database, with the user designated as its owner, (3) An inheritance relationship is established between the user and the organization and (4) The user is assigned all permissions within his organization. Steps (3) and (4) are carried out within the authorization service. These calls are implemented asynchronously to avoid response delays.

The distributed NoSQL database utilized for storing accounts and permissions does not feature builtin transaction management. This necessitates a careful approach to handle data consistency given the absence of transactional support. Given that several of previously described operations include separate database queries and the project employs Cassandra [19] database that does not support ACID transactions, maintaining consistency in the event of a failure in any step involved implementing a programmatic rollback.

### 4.2    Authentication and authorization flow

The authentication process begins with the user submitting credentials. Upon successful verification of these credentials by the TTP, the user receives an expiration token in response. The user is responsible for securely storing this token, and it is added as a Bearer token in subsequent requests for various tasks. Prior to executing each request, proper authorization is mandatory. To facilitate authorization, the user is obligated to transmit a login token for verification. Since each request is passed through API gateway, this component trancodes HTTP to gRPC request and calls *VerifyToken* for token validation. If the token is still valid, the authentication service will communicate with the authorization service (policy engine) to obtain granted user permissions. These permissions are then incorporated as a payload in a JSON Web Token (JWT). The JWT is appended to the initial request and shared exclusively among services within the distributed cloud. Each service is responsible for extracting permissions from the JWT.

The structure of the payload containing permissions is predetermined and comprises a username along with an array of permissions. Each element within the array is composed of the permission name, the permission kind (indicating whether it is ALLOW or DENY), and the permission ID. These components are separated by a delimiter. Based on the information within the JWT, if the user is authorized for the specific action in the service, the request proceeds to execution. On the contrary, if authorization is not granted, the process is terminated, and the user receives an appropriate message along with a status code.

Figure 2. shows a sequence diagram displaying a flow for creating new configuration by calling *PutStandaloneConfig* method. After token verification, user will get permissions related to his request. If needed permissions are present, request will be forwarded from API gateway to desired constellations service. If permission check fails and the JWT is not obtained, the user is immediatelly refused access.

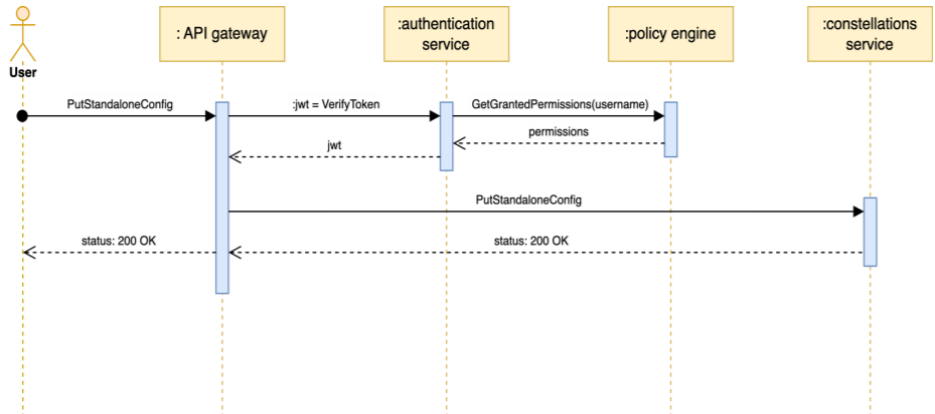

**Fig. 2.** Authorization flow for adding new configuration to the constellations platform

## 5    Discussion

In order to evaluate performance of the proposed solution, we considered two remarkable features for cloud environment: latency and throughput. These paramters are chosen based on the premise that centralized IAM server would reduce network overhead, provide lower latency and high availability. Latency is measured as total end-to-end wait time per request, seen by the user. Throughput is measured on server-side as the number of messages a server processes during a specific period of time. Two approaches were compared by performing *PutStandaloneConfig* request, based on these metrics. First approach measures latency and throught per request sent from client to constellation service through API gateway, following the flow shown on fig. 2. Second approach does not include immediate authorization. It redirects request directly to destination service, which is then responsible for calling policy engine in order to authorize request. Notably, most requests rely on multiple services, so the process of calling policy engine is repeated each time one service communicates with another. The results are shown in table 1. Based on the results, throughput is calculated as total number of requests sent / time period. Based on the results, throughput is ~1000req/min. It should be considered that the throughput is influenced not only by the volume of requests but also by the complexity of each request and the time required to process them.

Results show non-linear increase of latency as the number of requests icrease. It is arguable that as the communication among services increase for specific request, the latency is higher. Additionally, when the user is not authorized, the response in first approach is ~10ms, while it is ~30ms when services need to call policy engine. This also points out that proposed solution could be optimized by time-based or event-based token generation, which would remove calls to policy engine if there were no changes for user permissions.

**Table 1.** Performance results

| Number of requests | Flow with IAM server | Flow without IAM server |
|---|---|---|
| 10 | 2432ms | 2433ms |
| 500 | 21565ms | 22207ms |
| 1000 | 1m 33s | 2m 10s |

## 6  Conclusion

The primary objective of this paper is to showcase the implementation of a token-based authentication and authorization model, designed as an integral component of the authentication infrastructure for a distributed cloud. The paper introduced different elements of the Identity and Access Management (IAM) server. Token-based authentication and the secure storage of data are both developed relying on the services of the trusted third-party provider. However, TTP could not solve the issue of frequent user permission changes reflected in the token payload. For such cases, this paper proposed a novel approach of decoupling authentication and authorization tokens. The basic authorization request flow is outlined on an example request for adding configuration, while the IAM server is evaluated as an integrated part of a distributed cloud platform. The evaluation showed that IAM components ensures secure authentication and access control while upholding availability, scalability, and resilience. In future work, the IAM server will undergo further development and testing for in-teroperability in diverse projects. Furthermore, comprehensive research will be conducted on the viability of federated identity management as an improvement to the current solution.

**Acknowledgment.** Funded by the European Union (TaRDIS, 101093006). Views and opinions expressed are however those of the author(s) only and do not necessarily reflect those of the European Union. Neither the European Union nor the granting authority can be held responsible for them.